\documentclass[twocolumn,showpacs,superscriptaddress,preprintnumbers,amsmath,amssymb]{revtex4}
\usepackage{graphicx}
\usepackage{dcolumn}
\usepackage{bm}

\def\simge{\mathrel{
     \rlap{\raise 0.511ex \hbox{$>$}}{\lower 0.511ex \hbox{$\sim$}}}}
\def\simle{\mathrel{
     \rlap{\raise 0.511ex \hbox{$<$}}{\lower 0.511ex \hbox{$\sim$}}}}

\begin{document}
\preprint{SPhT-T07/028}
\title{Elliptic flow in transport theory and hydrodynamics}
\author{Cl\'ement Gombeaud}
\author{Jean-Yves Ollitrault}
\affiliation{Service de Physique Th\'eorique, CEA/DSM/SPhT,
  CNRS/MPPU/URA2306\\ CEA Saclay, F-91191 Gif-sur-Yvette Cedex.}

\date{\today}
\begin{abstract}
We present a new direct simulation Monte-Carlo method for solving the
relativistic Boltzmann equation. 
We solve numerically the 2-dimensional Boltzmann equation using this
new algorithm. We find that elliptic flow from this transport
calculation smoothly converges towards
the value from ideal hydrodynamics as the number of collisions per
particle increases, as expected on general theoretical grounds, but in
contrast with previous transport calculations. 
\end{abstract}

\maketitle
Ultrarelativistic nucleus-nucleus collisions at the Relativistic
Heavy Ion Collider (RHIC) have been argued to create a ``perfect
liquid'', with an extremely low viscosity~\cite{Tannenbaum:2006ch}. 
The essential piece of evidence is the large magnitude of elliptic 
flow at RHIC~\cite{Ackermann:2000tr}, which is as large as 
predicted by ideal-fluid models (which assume zero viscosity). 
Elliptic flow is an azimuthal asymmetry in the momentum distribution
of particles, projected onto the plane transverse to the beam
direction ($z$ axis): in a collision between two nuclei with non-zero impact
parameter, more particles are emitted parallel to impact 
parameter ($x$ axis) than perpendicular to it ($y$ axis). 
This asymmetry results from the almond shape of the 
overlap region between the colliding nuclei, which is transformed into
a momentum asymmetry by pressure gradients~\cite{Ollitrault:1992bk}:
microscopically, elliptic flow results from the interactions between
the produced particles, and is therefore a key observable of the dense
matter produced at RHIC.

In the microscopic language of particle physics, ideal fluid and low
viscosity translate into small mean free path of a particle between
two collisions or, equivalently, large rescattering cross sections
between the ``partons'' created in a collision. The description of the
system in 
terms of partons is itself questionable at the early, dense stage of
the collision, but it is nevertheless a helpful, intuitive picture. 
The natural question which arises then is: how large must the partonic
cross section be in order to achieve ideal-fluid behavior, i.e., local thermal
equilibirum?
How many partonic collisions are needed?

In this paper, we address this issue by solving
numerically a relativistic Boltzmann equation, and comparing the
results with relativistic hydrodynamics. 
It is well known that the Boltzmann equation reduces to hydrodynamics
when the mean free path is small (see~\cite{Marle} for a rigorous proof
in the relativistic case). 
Numerically, however, it has been found~\cite{Molnar:2004yh} that
hydrodynamics produces larger elliptic flow (by 30-40\%) than the
Boltzmann equation. In this paper, we address this issue using a
different method. A possible explanation for the discrepancy found 
in~\cite{Molnar:2004yh} is suggested at the end of this paper. 

The primary limitation of the Boltzmann equation is that it 
only applies to a dilute system, where the mean free path of a
particle is much larger than the distance between particles, so that 
one need only consider two-body collisions, many-body collisions
occurring at a much lower rate. 
There is no reason to believe that the RHIC liquid is dilute: 
interactions are nonperturbative, so that both the mean free path and
the distance between particles are of order $1/T$, where $T$ is the
temperature. 
Clearly, the Boltzmann equation cannot be used to directly 
simulate a heavy-ion collision; nevertheless, it has the potential of
giving us a grasp on deviations to thermalization. 

The relativistic formula for the collision rate between two beams of
particle densities $n_1$ and $n_2$, and arbitrary velocities ${\bf
  v_1}$ and ${\bf v_2}$ can be written as~\cite{Landau}: 
\begin{equation}
\label{relativistic}
\frac{dN_{\rm coll}}{dt d^3 {\bf x}}=\sigma n_1 n_2
\sqrt{|{\bf v_1}-{\bf v_2}|^2-|{\bf v_1}\times{\bf v_2}|^2/c^2},
\end{equation}
where $\sigma$ is the scattering cross section. 
There are several methods for implementing this collision rate in
Monte-Carlo algorithms. The most widely used method 
\cite{Zhang:1999bd,Molnar:2001ux} is the ZPC 
algorithm~\cite{Zhang:1997ej}: 
this algorithm treats particles as hard spheres, which
collide when their distance in the center-of-mass frame is smaller than
$r\equiv\sqrt{\sigma/\pi}$.
An alternative method is the stochastic collision algorithm proposed
by Xu and Greiner~\cite{Xu:2004mz}. 

Here, we implement a new algorithm, where relativistic effects are
incorporated in a physically transparent way. 
The second term under the square root in 
Eq.~(\ref{relativistic}) is the relativistic correction. It is 
simply understood, in the case of colliding hard spheres, as a 
geometrical effect resulting from the Lorentz contraction of the 
spheres. We take this contraction into account, in the Monte-Carlo 
simulation, by replacing spheres with oblate spheroids, whose polar 
axis is the direction of motion. 
As with the other algorithms, Lorentz covariance and locality 
are broken in the sense that collisions occur instantaneously at 
a finite distance. However, these violations are small if the 
system is dilute~\cite{Molnar:2001ux}, which is required by the
Boltzmann equation. The difference with the ZPC algorithm is
that the collision time is determined directly in the laboratory
frame, not in the center-of-mass frame. 

For sake of simplicity, we consider a two-dimensional gas of massless
particles in the transverse plane $(x,y)$. We thus neglect an
important feature of the dynamics of heavy-ion collisions, the fast
longitudinal expansion. As will be shown below, this is 
not crucial for elliptic flow, which is essentially unaffected by the
longitudinal expansion. Initial conditions for the $N$ particles are
generated randomly according to a gaussian distribution in coordinate
space, and an isotropic, thermal distribution in momentum space: 
\begin{equation}
\frac{dN}{d^2{\bf x}d^2{\bf p}}=\frac{N}{4\pi^2R_x R_y T^2}\exp\left(-
\frac{x^2}{2R_x^2}-\frac{y^2}{2 R_y^2}-\frac{p_t}{T(x,y)}\right),
\end{equation}
with $ R_y>R_x$ and $p_t\equiv\sqrt{p_x^2+p_y^2}$. The
temperature $T(x,y)$ is determined as a function of the local particle 
density per unit area, $n_{2d}$ according to $n_{2d}\propto T^2$, the
equation of state of a massless ideal gas in 2 dimensions.

\begin{figure}
\includegraphics*[width=0.4\linewidth]{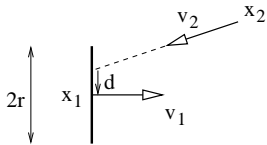}
\caption{Picture of a collision between a particle 1 of size $\sigma_{2d}=2 r$ and
  a pointlike particle $2$. The impact parameter $d$, as defined in
  Eq.~(\ref{defimpact}), is negative.
\label{fig:impact}}
\end{figure}

Particles interact via $2\to 2$ elastic collisions. 
We assume for simplicity that the total elastic  cross section 
$\sigma_{2d}$ is independent of the center-of-mass energy $s$. 
(In 2 dimensions, the cross section has the dimension of a length.) 
This is implemented in a Monte-Carlo calculation by treating each
particle as a rod  (due to Lorentz contraction) of length 
$r={1\over 2}\sigma_{2d}$ perpendicular to the velocity. Equivalently,
for each pair of colliding particles, one can assume that particle 1 has 
length $2 r$ and particle 2 is pointlike  (see Fig.~\ref{fig:impact}). 
The impact parameter of the collision is 
\begin{equation}
\label{defimpact}
d= -\frac{{\bf e_z}\cdot(({\bf x_2}-{\bf x_1})\times({\bf
    v_2}-{\bf v_1})/c)}{1-{\bf v_1}\cdot{\bf v_2}/c^2}
\end{equation}
This quantity is Lorentz invariant. 
Our collision algorithm is deterministic: the 
scattering angle in the center-of-mass frame $\theta^*$ is determined 
as a function of $d$, as in classical mechanics. 
The results presented here are obtained with an isotropic differential
cross section, i.e., $\theta^*= \pi(1-d/r)$. This prescription ensures
that particles move away from each other after the collision and do not
collide again.   

We now define two dimensionless numbers relevant to this problem. 
The average particle density per unit surface
$n_{2d}$ is~\footnote{Note that the surface thus 
  defined is a factor of 2 larger than in Ref.~\cite{Bhalerao:2005mm}
and a factor of 4 larger than in Ref.~\cite{Voloshin:1999gs}. This new
prescription gives the correct result for a uniform density profile,
as pointed out by A.~Poskanzer (private communication). The average
particle densities quoted in this paper are thus lower by a factor of 2
than in Tables I and II of Ref.~\cite{Bhalerao:2005mm}.}
\begin{equation}
n_{2d}=\frac{N}{4\pi R_x R_y}.
\end{equation}
The typical distance between two particles is $n_{2d}^{-1/2}$, while 
The mean free path of a particle between two collisions is  
$\lambda=1/(\sigma_{2d} n_{2d})$. The ratio of these two lengths is 
the dilution parameter $D$:
\begin{equation}
\label{defD}
D\equiv \frac{n_{2d}^{-1/2}}{\lambda}=\sigma_{2d} n_{2d}^{1/2}.
\end{equation}
As explained above, applicability of Boltzmann theory requires $D\ll
1$. 
In addition, locality and covariance require that the 
interaction length be much smaller than the mean free 
path~\cite{Molnar:2001ux,Zhang:1997ej}. In two dimensions, the
interaction length is $\sigma_{2d}$, and $\sigma_{2d}/\lambda=D^2$:
locality and covariance are thus recovered in the limit $D\ll 1$. 

The second dimensionless number is the Knudsen number ${\rm Kn}$, which
characterizes the degree of equilibration by comparing $\lambda$ with
the system size $R$. 
The latter quantity can be measured by any average of $ R_x$ and
$ R_y$. A natural choice for elliptic flow is~\cite{Bhalerao:2005mm}  
\begin{equation}
\label{defR}
R\equiv \left(\frac{1}{ R_x^2}+\frac{1}{ R_y^2}\right)^{-1/2}.
\end{equation}
The Knudsen number is then defined as 
\begin{equation}
\label{defK}
{\rm Kn}\equiv \frac{\lambda}{R}=\frac{1}{\sigma_{2d} n_{2d} R}.
\end{equation}
Hydrodynamics is the limit 
${\rm Kn}\ll 1$, while the limit ${\rm Kn}\gg 1$ corresponds to free-streaming 
particles. The values of $D$ and ${\rm Kn}$ can be tuned by varying 
the cross section $\sigma_{2d}$ and the number of particles $N$. 

Note that the mean free path is a local quantity, which depends on
space-time coordinates: in particular, it increases as the system
expands. Strictly speaking, $D$ and ${\rm Kn}$ defined by
Eqs.~(\ref{defD}) and (\ref{defK}) are initial values. Dimensional
analysis suggests that they are the relevant control parameters for
this problem.  

Solving the Boltzmann equation in the hydrodynamic limit requires both
$D\ll 1$ and ${\rm Kn}\ll 1$. This in turn requires a huge number of
particles $N$ in the 
Monte-Carlo simulation: inverting the above equations, one obtains
(assuming $ R_x\simeq  R_y$ for simplicity)
\begin{equation}
\label{npart2d}
N\simeq \frac{8\pi}{D^2 {\rm Kn}^2}.
\end{equation}
Since the computing time scales grows with $N$ like $N^{3/2}$, one cannot 
implement arbitrarily small values of $D$ and ${\rm Kn}$.
Instead, we choose to study numerically the dependence of elliptic
flow on $D$ and ${\rm Kn}$, and extrapolate to the hydro limit
$D={\rm Kn}=0$.

\begin{figure}
\includegraphics*[width=\linewidth]{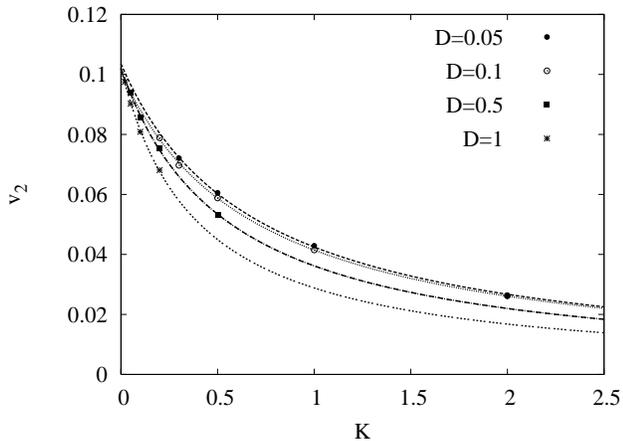}
\caption{Variation of the elliptic flow $v_2$ with the Knudsen number, 
  ${\rm Kn}$, for several values of the dilution parameter $D$. The
  statistical error on each point is $\delta v_2=7\times 10^{-4}$. 
  For each value of $D$, Monte-Carlo results are fitted using
  Eq.~(\ref{simpleformula}).  
\label{fig:v2vsK}}
\end{figure}

Fig.~\ref{fig:v2vsK} presents our results for the average elliptic 
flow $v_2\equiv \langle\cos 2\phi\rangle=\langle
(p_x^2-p_y^2)/(p_x^2+p_y^2)\rangle$. The aspect ratio of the initial
distribution is $ R_y=1.5  R_x$, and the Monte-Carlo
simulation has been pushed to very large times ($t=100 R$), so that
all collisions are taken into account.
Quite naturally, $v_2$ decreases monotonically to 0 as ${\rm Kn}$ increases
towards the free-streaming limit ${\rm Kn}\to +\infty$. 
For a given value of $D$, the variation with ${\rm Kn}$ is very well
described by the simple formula proposed in~\cite{Bhalerao:2005mm}:
\begin{equation}
\label{simpleformula}
v_2=\frac{v_2^{h}}{1+{\rm Kn}/{\rm Kn}_0},
\end{equation}
where the parameters $v_2^{h}$ and ${\rm Kn}_0$ are fit to our
Monte-Carlo results. 
$v_2^{h}$ is the limiting value of $v_2$ when ${\rm Kn}\to 0$,
expected to coincide with $v_2$ from hydrodynamics in the limit 
$D\to 0$. 
Surprisingly, the value of $v_2^h$ depends very little on $D$:
$v_2^h=0.102\pm 0.003$. This means that 
the hydrodynamic limit is more general than the Boltzmann equation 
and applies even if the system is not dilute. 
Unlike $v_2^h$, the parameter  ${\rm Kn}_0$ strongly depends on
$D$. We do not have a simple explanation for this dependence. 
However, only the limit $D\ll 1$ has a well-defined physical
interpretation, as it corresponds to the Boltzmann equation. For
larger values of $D$, locality and causality are broken, and the
physical interpretation of the results is less clear. 
For $D\ll 1$, our fit gives ${\rm Kn}_0=0.70\pm 0.03$.

An independent hydro calculation with the same initial conditions 
was done using the same code as  
in Ref.~\cite{Ollitrault:1992bk}. 
For sake of consistency~\cite{Molnar:2004yh},
the equation of state of the fluid is that of a 
two-dimensional ideal gas (i.e., the equation of state of a dilute gas, as
modeled by the Boltzmann equation), whose velocity of sound is 
$c_s=c/\sqrt{2}$~\cite{Heinz:2002rs}. The average $v_2$ at $t=100 R$ is 
$v_2^{\rm hydro}=0.101\pm 0.003$~(the error bar in the 
  hydro calculation is due to the fact that the calculation is pushed
  to very large times). This is compatible with the value $v_2^h$
  obtained from the Boltzmann calculation: Boltzmann transport theory
  and ideal hydrodynamics agree in the limit ${\rm Kn}\to 0$, as
  expected. 

The Knudsen number is closely related to the average number of
collisions per particle $\bar n_{\rm coll}$. From the definition,
Eq.~(\ref{defK}), one expects that the 
number of collisions per particle is $\sim 1/{\rm Kn}$. 
The proportionality constant can be computed exactly for a gaussian
distribution in the limit ${\rm Kn}\gg 1$~\cite{Heiselberg:1998es},
which yields the relation 
\begin{equation}
\label{ncollpart}
\bar n_{\rm coll}\equiv\frac{2 N_{\rm coll}}{N}=\frac{4\sqrt{2}}
{\pi^{3/2}\sqrt{1+\epsilon}}K\!\left(\frac{2\epsilon}{1+\epsilon}\right)
{\rm Kn}^{-1}
\simeq \frac{1.6}{\rm Kn},
\end{equation}
where the factor 2 means that each collision involves 2 particles, 
$\epsilon=(R_y^2-R_x^2)/(R_y^2+R_x^2)$ is 
the initial eccentricity, and 
$K(x)$ is the complete elliptic integral of the first kind. 
We used this formula to check numerically that our algorithm produces
the right number of collisions for large ${\rm Kn}$. In addition, our
numerical results show that the product $\bar n_{\rm coll}{\rm Kn}$ is
remarkably constant for all values of ${\rm Kn}$, so that
Eq.~(\ref{ncollpart}) holds within a few percent. 
Using Eq.~(\ref{simpleformula}), this shows that
an average of $2.3$ (resp. $9.1$) collisions per particle are required 
to achieve 50\% (resp. 80\%) of the hydrodynamic ``limit'' on $v_2$. 

\begin{figure}
\includegraphics*[width=\linewidth]{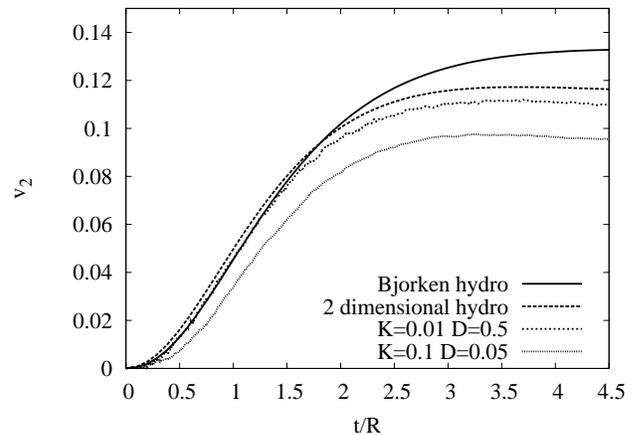}
\caption{Time dependence of the average elliptic flow $v_2$ from the
  transport model, from the corresponding 2-dimensional hydro
  calculation, and from usual 3-dimensional hydro with Bjorken
  longitudinal expansion.
\label{fig:v2vst}}
\end{figure}

Fig.~\ref{fig:v2vst} compares the time-dependence of elliptic flow in
hydro and in the transport model. The subtle point is that the
convergence of $v_2(t)$ towards hydro as ${\rm Kn}\to 0$ is not
uniform: for a given value of ${\rm Kn}$, deviations from hydro are
large at early times, and tend to decrease afterwards. More precisely,
it can be shown, following the same methods as in 
Ref.~\cite{Heiselberg:1998es}, that the early-time behavior is 
$v_2\propto t^3$ in the transport model, and $v_2\propto t^2$ in
hydro. This is compensated by the late-time behavior:
$v_2$ decreases slowly at large times in hydro (not seen in the
figure), while $v_2$ from the transport model stays constant after the
last collision has occurred.

Can our 2-dimensional results be used in the context of heavy-ion collisions? 
The essential difference in 3 dimensions is the fast longitudinal 
expansion due to the strong Lorentz contraction of the colliding
nuclei. 
Elliptic flow, however, is a purely transverse observable which is
little affected by the longitudinal expansion: 
Fig.~\ref{fig:v2vst} displays a comparison between the average
elliptic flow computed in 2-dimensional hydro and in 
3-dimensional hydro with Bjorken longitudinal
expansion (the initial time in this calculation is $\tau_0=R/4$,
corresponding to $\tau_0\simeq 0.4$~fm/c in a semicentral 
Au-Au collision)~\cite{Bjorken:1982qr}. Both yield similar results for the
magnitude and the time-dependence of $v_2$. 

One can reasonably expect that deviations from hydro are 
similar in 2 dimensions and in 3 dimensions. 
The number of collisions per particle, $\bar n_{\rm coll}$, is not the
right quantity for  carrying out the comparison (neither the transport
opacity~\cite{Molnar:2001ux}): 
in 3 dimensions, $\bar n_{\rm coll}$ is large at early times
(with a fixed partonic cross section, $\bar n_{\rm coll}$ diverges like
$\ln\tau_0^{-1}$ as the initial time $\tau_0$ goes to 0), but these 
collisions do not produce much elliptic flow (see Fig.~\ref{fig:v2vst}). 
As argued in Ref.~\cite{Bhalerao:2005mm}, the Knudsen number should be
evaluated at the time when elliptic flow develops $t\sim R/c_s$,
with $c_s\simeq 1/\sqrt{3}$ in the quark-gluon plasma phase. The mean
free path is $\lambda=1/\sigma n$ 
with $n=(1/ct)(1/S)(dN/dy)$ and $S=4\pi R_x R_y$, hence 
\begin{equation}
\label{Kexp}
\frac{1}{\rm Kn}=\sigma\frac{1}{S}\frac{dN}{dy} \frac{c_s}{c}.
\end{equation}
$dN/dy$ is the total (charged+neutral) multiplicity. For a
semi-central Au-Au collision at RHIC, $(1/S)(dN/dy)\simeq
0.8$~mb$^{-1}$. 
With this definition of ${\rm Kn}$, we expect that
Eq.~(\ref{simpleformula}) should hold approximately in 3 dimensions,
with ${\rm Kn}_0\sim 0.7$. 
Table~\ref{tab:sigma} gathers numerical estimates obtained 
using the same values of $\sigma$ as in Ref.\cite{Molnar:2004yh}. 
For $\sigma=20$~mb, corresponding to a QCD
cross-section of 47~mb, the transport result should be only $\simeq
10-15\%$ below hydro.   
\begin{table}[t]
\caption{Expected values of $v_2$ for semi-central Au-Au
  collisions at RHIC. For each value of the cross section, we quote
  the value of the equivalent Debye-screened, leading order QCD cross
  section \cite{Molnar:2004yh}.\label{tab:sigma}}  
\begin{tabular}{|c|c|c|c|}
\hline
Isotropic  $\sigma$ [mb] & Debye-screened $\sigma$ [mb]& ${\rm Kn}$ &
$v_2/{v_2^{\rm hydro}}$\\ \hline 
3 &7 & 0.72 & 0.49 \\ \hline
8& 19& 0.27& 0.72\\ \hline
20 & 47 & 0.11& 0.87\\ \hline
\end{tabular}
\end{table}

For this value of the cross section, Molnar and Huovinen 
find that $v_2$ from the transport calculation is lower by 30\% than
$v_2$ from hydro \cite{Molnar:2004yh}. Furthermore, the dependence of 
their results on $\sigma$ is at variance with our results. 
Using Eqs.~(\ref{simpleformula}) and (\ref{Kexp}), the deviations from
ideal hydro should decrease as $1/\sigma$ as $\sigma$ increases. 
This is very general: dissipative effects are expected to be 
linear in the viscosity $\eta$, which scales like $1/\sigma$. 
The discrepancy with hydro should be at least twice smaller with
$\sigma=47$~mb than with $\sigma=20$~mb, which is clearly not the case
in Fig.~1 of Ref.~\cite{Molnar:2004yh}. In our opinion, this
cannot be attributed to dissipative effects. 

The origin of the problem may be the dilution condition $D\ll 1$, 
which is not satisfied in previous transport calculations. 
The trick to reduce $D$ in the ZPC cascade algorithm is the ``parton
subdivision technique'': one multiplies the number of particles $N$ 
by a large number $l$, and one divides the partonic cross section
$\sigma$ by $l$, so that the Knudsen number ${\rm Kn}$ is 
unchanged. 
The distance between particles in three dimensions is $n^{-1/3}$,
which replaces $n_{2d}^{-1/2}$ in Eq.~(\ref{defD}). Taking parton
subdivision into account, one obtains
$D=\sigma n^{2/3}l^{-1/3}$.
The average particle density in a semicentral Au-Au collision at time
$t=R/c_s$ is $2.5$~fm$^{-3}$~\cite{Bhalerao:2005mm}. With
$\sigma=20$~mb and $l=180$~\cite{Molnar:2004yh}, $D\simeq 0.65$, which
is not very small compared to unity. The situation is worse at early
times due to the longitudinal expansion: $D\simeq 1.1$ at
$\tau_0=0.6$~fm/c. 
For ${\rm Kn}=0.11$, carrying out the calculation with $D\simeq 1$ 
leads to overestimate the deviation from hydro by at least a factor of
2 (see Fig.~\ref{fig:v2vsK}). 
The reason why previous calculations were done with such large values
of $D$ is that small values are hard to achieve 
numerically: in 3 dimensions, Eq.~(\ref{npart2d}) is replaced with 
$N\propto D^{-3}{\rm Kn}^{-3}$, and the computing time grows like 
$N^{4/3}\propto D^{-4}{\rm Kn}^{-4}$. This was our motivation for
studying first the 2-dimensional case.

In summary, we have implemented a new algorithm for solving the relativistic 
 Boltzmann equation.  We have studied the convergence of the relativistic 
 Boltzmann equation to relativistic hydrodynamics. Since this requires
 both a dilute system ($D\ll 1$) and a small mean free path (${\rm Kn}\ll 1$),
 which is costly in terms of computer time, we have restricted our
 study to two dimensions. This preliminary study has only addressed
 the average value of the elliptic flow; differential results as a
 function of $p_t$, as well as results on $v_4$, will be presented in
 a forthcoming publication. We have shown that the average elliptic flow
 computed in the transport algorithm converges smoothly towards the
 hydrodynamics value as the strength of final-state interactions
 increases. Isotropic partonic cross sections of 3~mb and 20~mb should
 produce respectively $\sim 50\%$ and $\sim 90\%$ of the hydro value for $v_2$. 

\section*{Acknowledgments}

We thank Carsten Greiner, Denes Molnar, Aihong Tang and Zhe Xu  for
useful comments on the manuscript.


\begin{thebibliography}{99}

\bibitem{Tannenbaum:2006ch}
  M.~J.~Tannenbaum,
  Rept.\ Prog.\ Phys.\  {\bf 69}, 2005 (2006).

\bibitem{Ackermann:2000tr}
  K.~H.~Ackermann {\it et al.}, 
  Phys.\ Rev.\ Lett.\  {\bf 86}, 402 (2001).

\bibitem{Ollitrault:1992bk}
  J.~Y.~Ollitrault,
  Phys.\ Rev.\  D {\bf 46}, 229 (1992).

\bibitem{Marle}
C. Marle, 
Annales Poincare Phys.Theor. {\bf 10},67 (1969).

\bibitem{Molnar:2004yh}
  D.~Molnar and P.~Huovinen,
  Phys.\ Rev.\ Lett.\  {\bf 94}, 012302 (2005).

\bibitem{Landau}
  L.~D.~Landau, E.~M.~Lifshitz,
{\it The Classical Theory of Fields\/}, 4$^{\rm th}$ Edition (Pergamon
  Press, 1975), page 36. 

\bibitem{Zhang:1999bd}
  B.~Zhang, C.~M.~Ko, B.~A.~Li and Z.~w.~Lin,
  Phys.\ Rev.\  C {\bf 61}, 067901 (2000);
  Z.~W.~Lin, C.~M.~Ko, B.~A.~Li, B.~Zhang and S.~Pal,
  Phys.\ Rev.\  C {\bf 72}, 064901 (2005).

\bibitem{Molnar:2001ux}
  D.~Molnar and M.~Gyulassy,
  Phys.\ Rev.\  C {\bf 62}, 054907 (2000);
  Nucl.\ Phys.\ A {\bf 697}, 495 (2002)
  [Erratum-ibid.\ A {\bf 703}, 893 (2002)].

\bibitem{Zhang:1997ej}
  B.~Zhang,
  Comput.\ Phys.\ Commun.\  {\bf 109}, 193 (1998).
  B.~Zhang, M.~Gyulassy and Y.~Pang,
  Phys.\ Rev.\ C {\bf 58}, 1175 (1998).

\bibitem{Xu:2004mz}
  Z.~Xu and C.~Greiner,
  Phys.\ Rev.\  C {\bf 71}, 064901 (2005).

\bibitem{Bhalerao:2005mm}
  R.~S.~Bhalerao, J.~P.~Blaizot, N.~Borghini and J.~Y.~Ollitrault,
  Phys.\ Lett.\  B {\bf 627}, 49 (2005).

\bibitem{Voloshin:1999gs}
  S.~A.~Voloshin and A.~M.~Poskanzer,
  Phys.\ Lett.\  B {\bf 474}, 27 (2000).

\bibitem{Heinz:2002rs}
  U.~W.~Heinz and S.~M.~H.~Wong,
  Phys.\ Rev.\  C {\bf 66}, 014907 (2002).

\bibitem{Heiselberg:1998es}
  H.~Heiselberg and A.~M.~Levy,
  Phys.\ Rev.\  C {\bf 59}, 2716 (1999).

\bibitem{Bjorken:1982qr}
  J.~D.~Bjorken,
  Phys.\ Rev.\  D {\bf 27}, 140 (1983).

\end{thebibliography}
\end{document}